\documentclass[10pt,aps,prl,preprintnumbers,superscriptaddress,showpacs,
twocolumn,floatfix]{revtex4-1}
\usepackage{graphicx}
\usepackage{amsmath}
\usepackage{slashed}

\newcommand{\gsim}{\;\rlap{\lower 3.5 pt \hbox{$\mathchar \sim$}} \raise 1pt
\hbox {$>$}\;}
\newcommand{\lsim}{\;\rlap{\lower 3.5 pt \hbox{$\mathchar \sim$}} \raise 1pt
\hbox {$<$}\;}

\begin{document}

\title{
\boldmath High-Energy Limit of Quantum Electrodynamics beyond Sudakov
Approximation
\unboldmath}
\author{Alexander A. Penin}
\affiliation{Department of Physics, University of Alberta, Edmonton, Alberta T6G
2J1, Canada}
\affiliation{Institut f\"ur Theoretische Teilchenphysik,
Karlsruhe Institute of Technology, 76128 Karlsruhe, Germany}
\begin{abstract}
We study the high-energy behavior of the scattering amplitudes in  quantum
electrodynamics beyond the leading order of the small electron mass expansion
in the leading logarithmic approximation. In contrast to the  Sudakov
logarithms, the mass-suppressed double-logarithmic radiative corrections  are
induced by a soft electron pair exchange and result in enhancement of the
power-suppressed contribution. Possible applications of our result to the
analysis of the high-energy processes in quantum chromodynamics is also
discussed.

\end{abstract}
\preprint{ALBERTA-THY-18-14}

\maketitle
In a renowned paper \cite{Sudakov:1954sw} V.V.~Sudakov derived the leading
asymptotic behavior of an electron scattering amplitude in quantum
electrodynamics (QED) at high energy. It is determined by the ``Sudakov''
radiative corrections, which include the second power of the  large logarithm of
the electron mass $m_e$ divided by a characteristic momentum transfer of the
process per each power of the fine structure constant $\alpha$. Sudakov double
logarithms exponentiate and result in a strong universal suppression of any
electron scattering amplitude with a fixed number of emitted photons in the
limit when all the kinematic invariants of the process are large. This result
plays a fundamental role in particle physics. Within different approaches it has
been  extended to the nonabelian gauge theories and to the subleading logarithms
\cite{Frenkel:1976bj,Mueller:1979ih,Collins:1980ih,Sen:1981sd,Sterman:1986aj,
Catani:1998bh}, which is crucial for a wide class of applications  from deep
inelastic scattering to Drell-Yan processes and the Higgs boson production. At
the same time no significant progress has been achieved in the study of the
logarithmically enhanced corrections to the subleading contributions suppressed
by a power of electron mass at high energies. However, the power-suppressed
contributions are of great interest. They can become asymptotically dominant at
very high energies due to Sudakov suppression of the leading  terms. At the
intermediate energies the power corrections in many cases are phenomenologically
important
\cite{Bernreuther:2004ih,Bernreuther:2005gw,Bonciani:2007eh,Bonciani:2008ep}.
Moreover, in contrast to the Euclidean operator product expansion
\cite{Wilson:1969zs} or nonrelativistic threshold dynamics \cite{Caswell:1985ui}
very little is known about the general all-order structure of the large
logarithms beyond the leading-power approximation in the high-energy limit,
which is a real challenge for the effective field theory approach. This problem
is now actively discussed in various contexts (see {\it e.g.}
\cite{Feucht:2004rp,Laenen:2010uz,Becher:2013iya,Banfi:2013eda,deFlorian:2014vta,
Anastasiou:2014lda}). In this Letter we make the first step toward the solution
of the problem and generalize the result of Ref.~\cite{Sudakov:1954sw} to the
leading power-suppressed contribution. We present a detailed analysis of the
electron scattering in the external field and later discuss the extension of the
result to more complex processes.

\begin{figure}[t]
\vspace*{-15pt}
\begin{center}
\hspace*{20pt}\includegraphics[width=7.5cm]{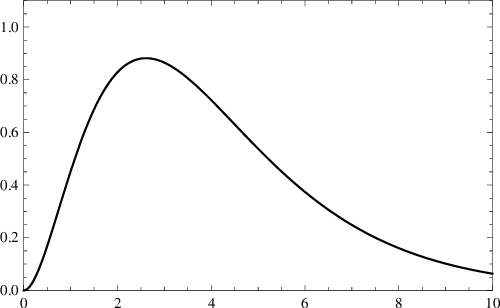}\\[0pt]
\hspace*{15pt}$x$\\
\vspace*{-92pt}
\hspace*{-220pt}$f(x)$
\vspace*{70pt}
\end{center}
\caption{\label{fig::fx}  The result of the numerical evaluation of the function
$f(x)=-3F^{(1)}_1$.
}\end{figure}

The  amplitude  ${\cal F}$  of the electron scattering in an external
field can be parameterized in the standard way  by the  Dirac and Pauli
form factors
\begin{equation}
{\cal F}=\bar{\psi}(p_2)\left(\gamma_\mu F_1+{i\sigma_{\mu\nu}q^\nu \over 2m_e}
F_2\right)\psi(p_1).
\label{eq::Dirac}
\end{equation}
The Pauli form factor $F_2$ does not contribute in the approximation discussed
in this Letter and we mainly focus on the high-energy behavior of the Dirac form
factor $F_1$. We consider the limit of the on-shell electron $p_1^2=p_2^2=m_e^2$
and the large Euclidean momentum transfer $Q^2=-(p_2-p_1)^2$ when the ratio
$\rho\equiv {m_e^2/Q^2}$ is positive and small. The Dirac form factor can then
be expanded in an asymptotic series in $\rho$
\begin{equation}
F_1=S_\lambda\sum_{n=0}^\infty \rho^n F^{(n)}_1,
\label{eq::F1}
\end{equation}
where $F^{(n)}_1$ are given by the power series in $\alpha$ with the
coefficients depending on $\rho$ only logarithmically. The factor
$S_\lambda=\exp{\left[-{\alpha\over 2\pi}B(\rho)\,
\ln\left({\lambda^2/m_e^2}\right)\right]}$ with $B(\rho)=\ln\rho+{\cal O}(1)$
accounts for the universal singular dependence of the amplitude on the auxiliary
photon mass $\lambda$ introduced to regulate the infrared divergences
\cite{Yennie:1961ad}. In the double-logarithmic approximation the leading term
is given by the Sudakov exponent $F^{(0)}_1=e^{-x}$, with $x={\alpha\over
4\pi}\ln^2\rho$ \cite{Jackiw:1968zz}. Let us outline our approach for the
analysis of the power-suppressed double-logarithmic contributions. We use the
expansion by regions method \cite{Beneke:1997zp,Smirnov:1997gx,Smirnov:2002pj}
to get a systematic expansion of the Feynman integrals in $\rho$. In this method
the coefficients $F^{(n)}_1$  are given by the sum over contributions of
different virtual momentum regions. Each contribution is represented by a
Feynman integral which in general is divergent. These spurious divergences
result from the process of  scale separation and have to be dimensionally
regulated. The singular terms cancel out in the sum of all regions but can be
used to determine the logarithmic contributions to $F^{(n)}_1$.  The
double-logarithmic contributions are determined by the leading singular behavior
of the integrals and can be found by the method developed in Refs.
\cite{Sudakov:1954sw,Jackiw:1968zz,Gorshkov:1966ht}. Though the method is blind
to the power corrections, it can be applied in this case since the expansion by
regions provides the integrals which are {\it homogeneous} in the expansion
parameter. Sudakov logarithms are produced by the soft virtual photons, which
are collinear to either $p_1$ or $p_2$.  We have found that such a configuration
of virtual momenta does not produce  double logarithms in the first order in
$\rho$. This observation agrees with the analysis \cite{Korchemsky:1987wg} of
the cusp anomalous dimension, which determines the double-logarithmic
corrections to the light-like Wilson line with a cusp. For the large cusp angle
corresponding to the limit $\rho\to 0$  from the result of
Ref.~\cite{Korchemsky:1987wg} one gets
\begin{equation}
\Gamma_{cusp}=-{\alpha\over\pi}\ln\rho\left(1+{\cal O}(\rho^2)\right),
\end{equation}
with vanishing first-order term in $\rho$. Nevertheless, the ${\cal O}(\rho)$
double-logarithmic contribution does exist but originates from completely
different virtual momentum configuration described  below. Let us consider an
electron propagator $S={\slashed{p}_i-\slashed{l}+m_e\over (p_i-l)^2-m_e^2}$,
where $l$ is the momentum of a virtual photon with the propagator
$D={-g_{\mu\nu}\over l^2-\lambda^2}$. In the soft-photon limit $l\to 0$ the
electron propagator becomes eikonal $S\approx-{\slashed{p}_i+m_e\over 2p_il}$
and develops  a collinear singularity when $l$ is parallel to $p_i$.
Alternatively, we may consider the soft-electron limit $l'\to 0$, where
$l'=p_i-l$. Then the electron propagator becomes scalar $S\approx{m_e\over
l'^2-m_e^2}$ while the photon propagator becomes eikonal
$D\approx{g_{\mu\nu}\over 2p_il'-m_e^2+\lambda^2}$. Thus the roles of the
electron and photon propagators are exchanged. The existence of non-Sudakov
double-logarithmic contributions due to soft electron exchange has actually been
known for a long time \cite{Gorshkov:1966ht,Fadin:1997sn}. However in our case
this virtual momentum configuration does not produce a  double-logarithmic
contribution in one loop because the  momentum shift distorts  the eikonal
structure of the second electron propagator and removes the soft singularity at
small $l'$ necessary to get  the second power of the large logarithm.
\begin{figure}[t]
\includegraphics[width=1.5cm]{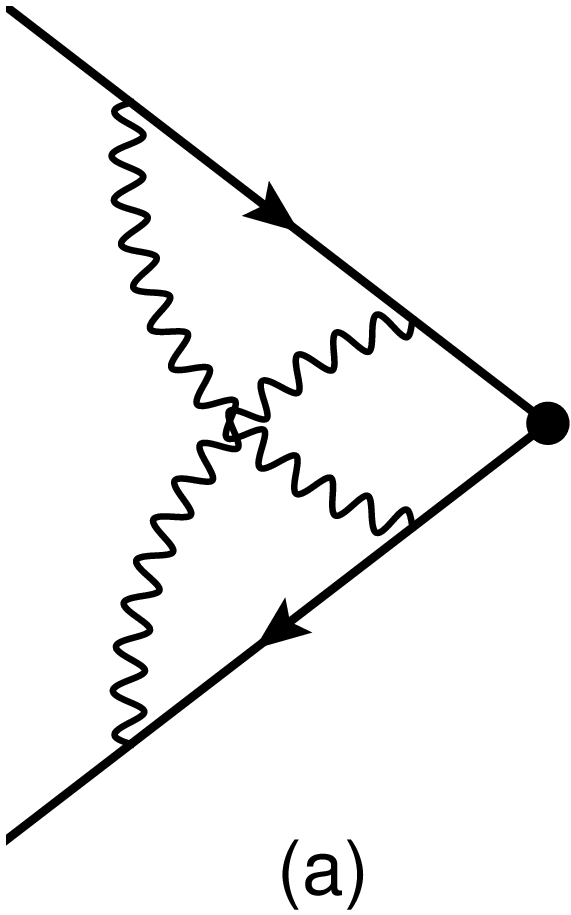}
\hspace*{10mm}\includegraphics[width=1.5cm]{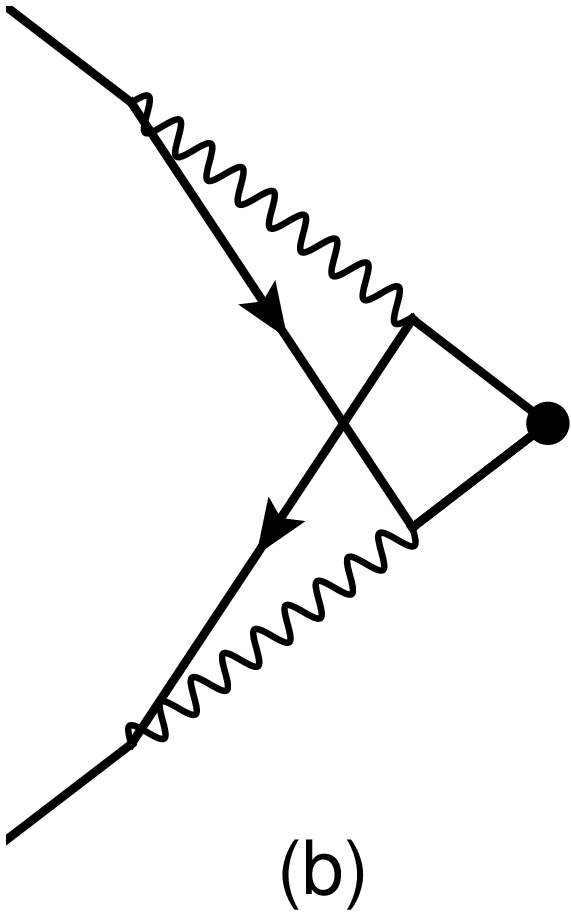}
\hspace*{10mm}\includegraphics[width=1.5cm]{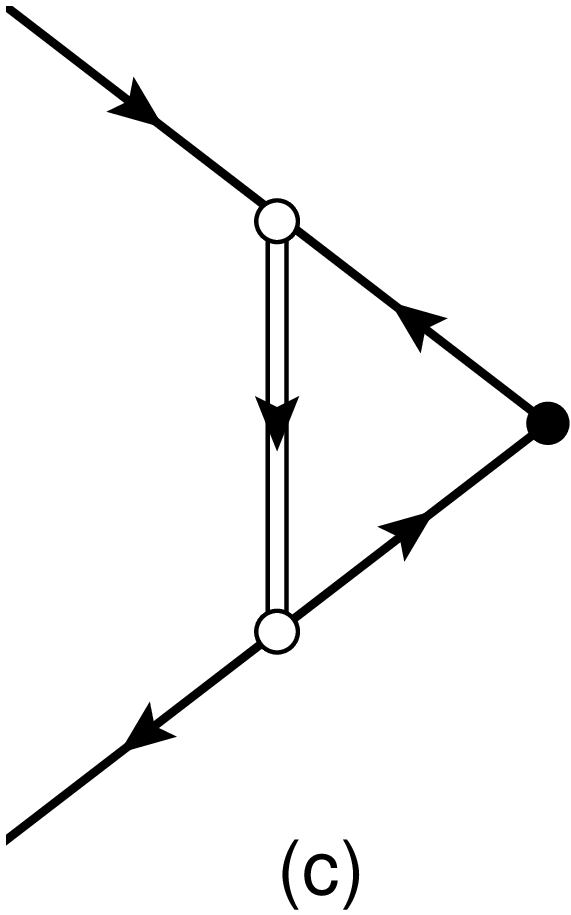}
\caption{\label{fig::diagram2loop}  Different representations of the
two-loop Feynman diagram giving the leading power-suppressed double-logarithmic
contribution. In figure (c) the double line arrow represents the soft electron
pair propagator and the empty blobs represent  the nonlocal interaction of the
soft electron pair to the eikonal electrons  and positrons.
}
\end{figure}
This may be avoided only in the two-loop diagram  of  nonplanar topology,
Fig.~\ref{fig::diagram2loop}(a). After shifting the photon virtual momenta by
$p_1$ and $p_2$  the diagram can be twisted into the shape of
Fig.~\ref{fig::diagram2loop}(b,c) with soft  electron pair exchange between the
eikonal lines.  The corresponding contribution has an explicit suppression
factor $m_e^2$ from two soft electron propagators. Hence the integration over
the virtual momenta can be performed in the leading order of the small electron
mass expansion. Note that in the case under consideration the electron mass
regulates both soft and collinear  divergences and we can put $\lambda=0$.  The
calculation is conveniently performed by using the light-cone coordinates where
$p_1\approx {p_1}_-$ and $p_{2}\approx{p_2}_+$.  In this representation only the
interaction of the transverse photons to soft electrons is not mass-suppressed
and  we can use ${g^\perp_{kl}\over 2p_il}$ for the eikonal photon propagators.
To get the double-logarithmic part of the correction we use  Sudakov
parametrization  of a virtual momentum $l=up_1+vp_2+l_\perp$. After
integrating over the transverse components $l_\perp$ we get the following
representation of the two-loop power-suppressed form factor
\begin{equation}
\left.F_1^{(1)}\right|_{2-loop}=-4x^2\!\!\int\!\! K(\eta_1,\eta_2,\xi_1,\xi_2)
 {\rm d}\eta_1{\rm d}\eta_2{\rm d}\xi_1{\rm d}\xi_2,
\label{eq::F12loop}
\end{equation}
where  $\eta=\ln v/\ln\rho $, $\xi=\ln u/\ln\rho$,  the integration goes over
the four-dimensional cube $0<\eta_i,~\xi_i<1$, and the kernel
\begin{eqnarray}
 K(\eta_1,\eta_2,\xi_1,\xi_2)&=&
 \theta(1-\eta_1-\xi_1)\theta(1-\eta_2-\xi_2)
 \nonumber\\
 && \times\theta(\eta_2-\eta_1) \theta(\xi_1-\xi_2)
 \label{eq::kernel2loop}
\end{eqnarray}
selects the kinematically allowed region of double-logarithmic integration. This
gives $F^{(1)}_1=-{x^2\over 3}+{\cal O}(x^3)$, in agreement with
\cite{Bernreuther:2004ih,Mastrolia:2003yz}.
\begin{figure}[t]
\includegraphics[width=1.5cm]{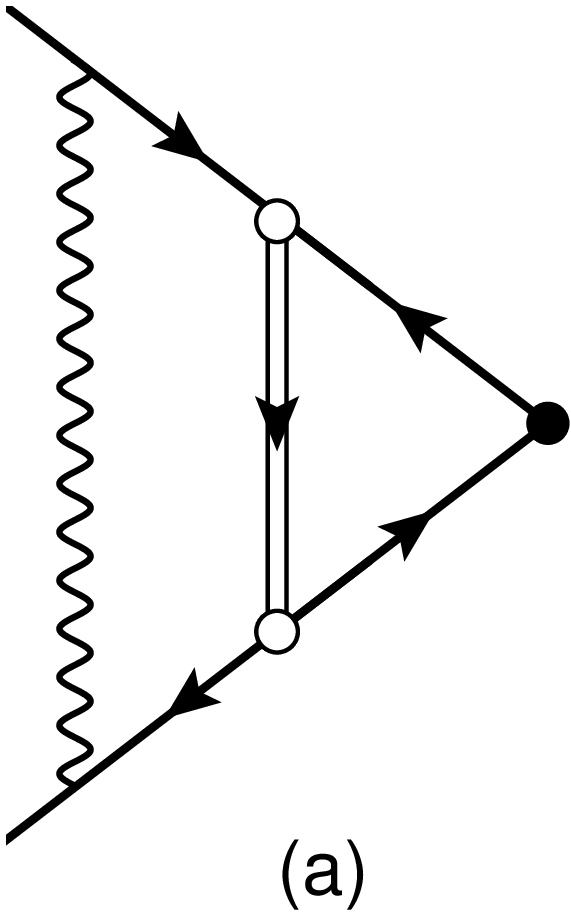}
\hspace*{10mm}\includegraphics[width=1.5cm]{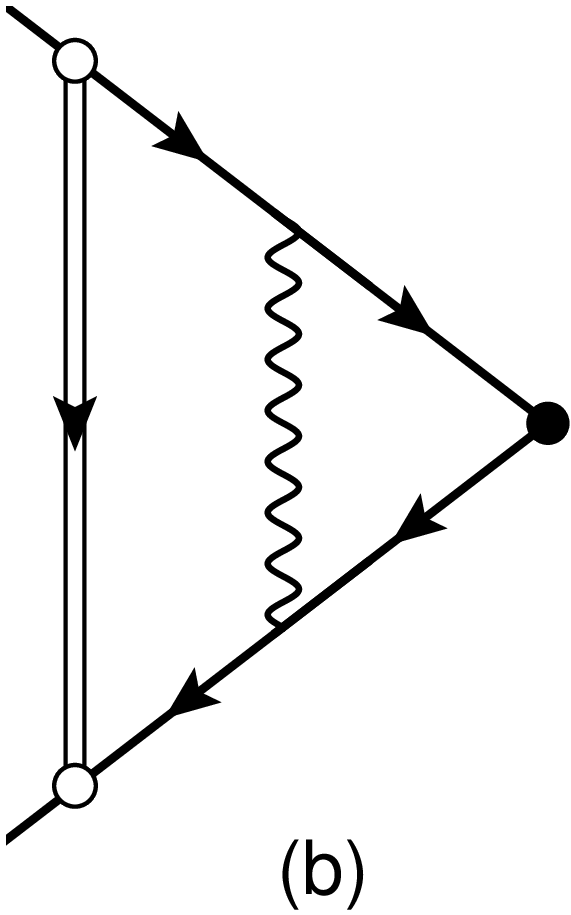}\hspace*{10mm}
\includegraphics[width=1.5cm]{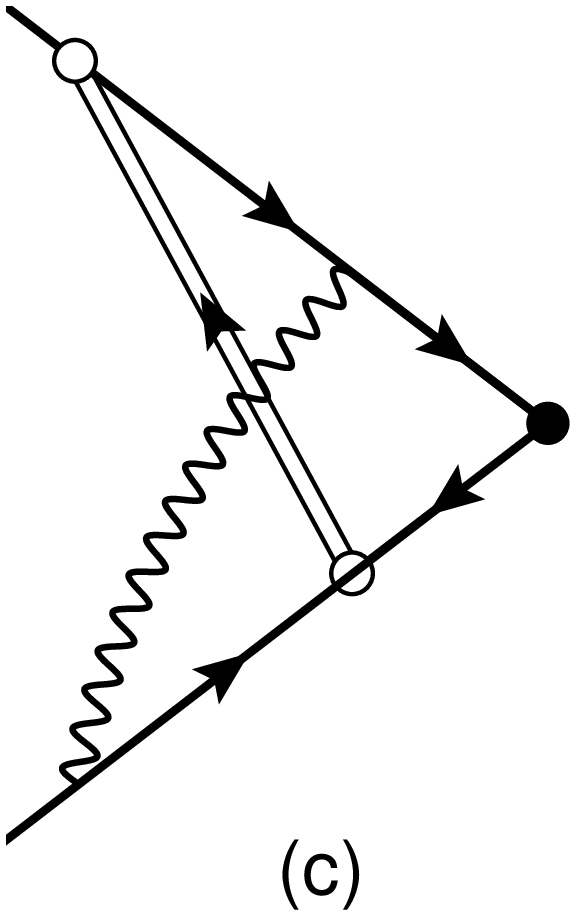}\\[2mm] \includegraphics[width=1.5cm]{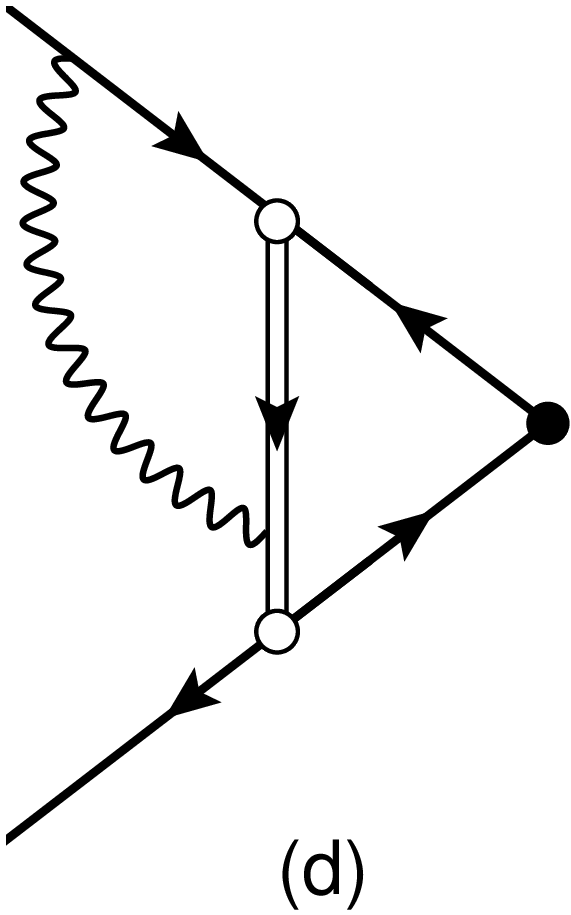}
\hspace*{10mm}\includegraphics[width=1.5cm]{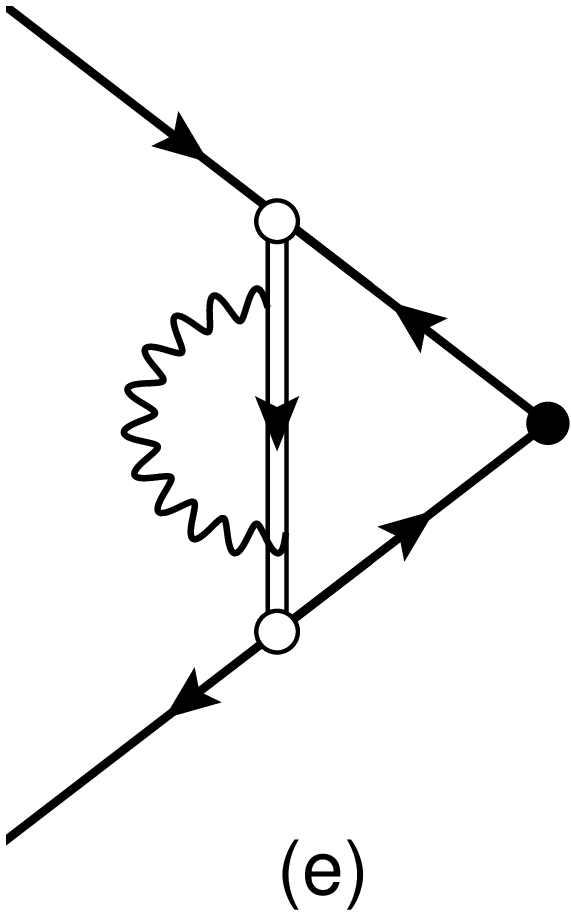}
\caption{\label{fig::diagram3loop} Feynman diagrams contributing to the
double-logarithmic correction factors $\phi^{a,b,c,d,e}$, Eq.~(\ref{eq::phi}).
}
\end{figure}
The higher-order double-logarithmic corrections are generated in a usual way
through the exchange of soft photons with the propagator ${-g_{+-}\over
l^2-\lambda^2}$.  The relevant topologies of the three-loop diagrams   are given
in Fig.~\ref{fig::diagram3loop}. Note that the soft photon exchange between the
soft and internal eikonal electron lines does not produce the
double-logarithmic correction. By using the factorization properties of the soft
photon contribution \cite{Sudakov:1954sw} we find the following representation
of the all-order double-logarithmic result
\begin{eqnarray}
 F^{(1)}_1&=&-4x^2\int {\phi}^a(\eta_2,\xi_1) {\phi}^b(\eta_1,\xi_2)
 {\phi}^c(\eta_1,\xi_1){\phi}^c(\xi_2,\eta_2)
\nonumber  \\
 &&
 \times
{\phi}^d(\eta_1,\xi_1) {\phi}^d(\xi_2,\eta_2)
{\phi}^d(\eta_2,\xi_1) {\phi}^d(\xi_1,\eta_2)  \nonumber  \\
 &&
 \times  {\phi}^e(\eta_1,\eta_2,\xi_1,\xi_2) K(\eta_1,\eta_2,\xi_1,\xi_2)\,
 {\rm d}\eta_1{\rm d}\eta_2{\rm d}\xi_1{\rm d}\xi_2
 \nonumber\\
&=&-4x^2\int \exp{\big[-x\left(1-2\eta_1\xi_1+4\eta_1\xi_2-2\eta_2\xi_2\right)\big]}
\nonumber  \\
&&
\times
K(\eta_1,\eta_2,\xi_1,\xi_2)\,{\rm d}\eta_1{\rm d}\eta_2{\rm d}\xi_1{\rm d}\xi_2,
\label{eq::F1integral}
 \end{eqnarray}
where the Sudakov correction factors corresponding to
Fig.~\ref{fig::diagram3loop}(a-e) are
\begin{eqnarray}
{\phi}^a(\eta,\xi)&=&\exp{\big[-x\left(1-2(\eta+\xi)-(\eta-\xi)^2\right)\big]},
\nonumber\\
{\phi}^b(\eta,\xi)&=&\exp{\big[-2x\eta\xi\big]},
\nonumber\\
{\phi}^c(\eta,\xi)&=&\exp{\big[x\eta\left(2+\eta-2\xi\right)\big]},
\nonumber\\
{\phi}^d(\eta,\xi)&=&\left[{\phi}^c(\eta,\xi)\right]^{-1},
\nonumber\\
{\phi}^e(\eta_i,\eta_j,\xi_i,\xi_j)&=&
\exp{\big[2x\left(\eta_i-\eta_j\right)\left(\xi_i-\xi_j\right)\big]},
\label{eq::phi}
\end{eqnarray}
respectively. We are not able to find the result for the  four-fold
integral~(\ref{eq::F1integral}) in a closed analytic form. However,  the
coefficients of the series
\begin{table}[t]
  \begin{ruledtabular}
    \begin{tabular}{c|c|c|c|c|c|c|c}
      $n$ & $1$ & $2$ & $3$ & $4$ & $5$ & $6$ & $7$
      \\
      \hline
      \hline
      $(-1)^nn!c_n$ &  ${4\over 5}$ & ${137\over  210}$ & ${341\over 630}$
      & ${23704\over 51975}$ & ${10529\over 27027}$ &
      ${96553\over 286650}$ & ${1352489\over 4594590}$
      \\
    \end{tabular}
    \end{ruledtabular}
    \caption{\label{tab::cn}
      The   normalized coefficients of the series~(\ref{eq::F1series}) up to
      $n=7$.}
\end{table}
\begin{equation}
F^{(1)}_1=-{x^2\over 3}\left(1+\sum_{n=1}^\infty c_nx^n\right)
\label{eq::F1series}
\end{equation}
can in principle be analytically computed for any given $n$ corresponding to the 
$(n+2)$-loop double-logarithmic contribution.  The first seven coefficients of 
the series are listed in Table~\ref{tab::cn}. The series~(\ref{eq::F1series}) is 
useful only for sufficiently small $x$.  For  $x\sim 1$ the 
integral~(\ref{eq::F1integral}) can be computed numerically. The result of the 
numerical evaluation is presented in Fig.~\ref{fig::fx} for the function 
$f(x)=-3F^{(1)}_1$. The function rapidly grows to the maximum value 
$f(x)=0.881566\ldots$ at $x=2.60904\ldots$ corresponding to
\begin{equation}
F^{(1)}_1=-0.293855\ldots
\label{eq::F1asymptotic}
\end{equation}
and then decays monotonically. For  large $x\gg 1$ the asymptotic behaviour of 
the function is exponential $f(x)\sim e^{-{x\over 2}+2\ln x}$. However, due to 
the exponential suppression the double logarithmic result is not an accurate
approximation  already for $x\approx 10$  where  the subleading terms 
proportional to powers of $\alpha\ln\rho\approx 1/3$  have to be taken into 
account.

Unlike the  Sudakov double logarithms,  the  leading power-suppressed
double-logarithmic corrections depend not only on the charges of the initial and
final states but also on the details of the scattering process. For example, the
non-singlet ({\it i.e.} without a closed electron loop) ${\cal O}(\rho)$
double-logarithmic corrections to the scalar form factor vanish to all orders in
$\alpha$ due to a specific Lorentz and Dirac structure of the soft electron pair
interaction with the eikonal electrons and positrons. A less trivial example is
the Pauli form factor. The expansion of  $F_2$ in $\rho$ ({\it cf.}
Eq.~(\ref{eq::F1}))  starts with the first order term $F^{(1)}_2$. In the
double-logarithmic approximation $F^{(1)}_2=0$  and for the leading mass
correction from the soft electron pair exchange  we obtain
$F^{(2)}_2=4F^{(1)}_1$, in agreement with
\cite{Bernreuther:2004ih,Mastrolia:2003yz}. Thus the ${\cal O}(\rho)$
corrections are universally related to the soft electron pair exchange and can
be be obtained as a straightforward generalization of our analysis for more
complicated processes such as Bhabha scattering, where only the leading result
of the small electron mass expansion is available  in two loops
\cite{Penin:2005eh,Penin:2005kf}. Moreover, up to two loops the structure of the
${\cal O}(\rho)$ double-logarithmic correction in quantum chromodynamics (QCD)
is similar to the one in QED. In particular, the double-logarithmic
power-suppressed term in two-loop corrections to the heavy-quark vector form
factor  differs from the QED result only by the $C_F^2-C_AC_F/2$ color factor of
the diagram in Fig.~\ref{fig::diagram2loop}. Thus our method can be applied to
the calculation of the dominant two-loop power-suppressed corrections to the
high-energy processes involving heavy quarks. For the energies ranging from
approximately 10 to 100 times the  heavy-quark mass we have $|\ln\rho|\gg 1$ and
$\rho\ln^4\rho\sim 1$, {\it i.e.} the double-logarithmic terms saturate the
power-suppressed contribution and are comparable in magnitude  to the
nonlogarithmic leading-power corrections in the strong coupling constant,  which
are phenomenologically significant. Beyond the two-loop approximation our result
is not directly applicable to the QCD amplitudes since the eikonal gluons in
Fig.~\ref{fig::diagram2loop}(b) can radiate  soft gluons producing additional
double-logarithmic corrections.  As a consequence,  the leading power-suppressed
double-logarithmic corrections to the heavy-quark vector form factor get a
nonabelian contribution in every order of perturbation theory in contrast to the
purely abelian Sudakov double logarithms.

To summarize, we have generalized the result of Sudakov~\cite{Sudakov:1954sw} to 
the leading power-suppressed contribution.  This is an important step towards a 
systematic renormalization group analysis of the  high energy behavior of the 
gauge theory amplitudes beyond the leading power approximation.  The  leading 
power-suppressed double-logarithmic corrections reveal a few characteristic 
features which distinguish them from the Sudakov double logarithms. In 
particular, they  are induced by a soft electron pair exchange and result in an 
enhancement of the power-suppressed contribution in the region where the double 
logarithmic approximation is applicable. In QCD our method can be used for the 
analysis of the high-energy processes involving heavy quarks up to two loops. 
Extending the analysis to the higher orders of perturbative  QCD and to 
subleading logarithms is  a very interesting problem which is beyond the scope 
of this Letter.

\begin{acknowledgements}
I would like to thank R. Bonciani,  K. Melnikov,  V. Smirnov, and  N. Zerf
for  useful discussions and collaboration. This research  was supported in part
by NSERC and  Perimeter Institute for Theoretical Physics.
Research at Perimeter Institute is supported by the Government of Canada
through Industry Canada and by the Province of Ontario through the Ministry
of Research and Innovation.
\end{acknowledgements}


\end{document}